\documentclass[aps,prl,reprint,superscriptaddress]{revtex4-2}

\usepackage{amsmath}
\usepackage{amssymb}
\usepackage[utf8]{inputenc}
\usepackage{siunitx} 
\usepackage{xcolor}
\usepackage{natbib}
\usepackage{graphicx}
\usepackage{braket}
\usepackage{comment}
\usepackage{bm}
\usepackage{epstopdf}

\begin{document}

\title{Excitonic fine structure of epitaxial Cd(Se,Te) on ZnTe type-II quantum dots}

\date{\today}

\author{Petr Klenovsk\'y}
\email{klenovsky@physics.muni.cz}
     \affiliation{Department of Condensed Matter Physics, Faculty of Science, Masaryk University, Kotl\'a\v{r}sk\'a~267/2, 61137~Brno, Czech~Republic}
     \affiliation{Czech Metrology Institute, Okru\v{z}n\'i 31, 63800~Brno, Czech~Republic}

\author{Piotr Baranowski}
\email{baranowski@ifpan.edu.pl}
    \affiliation{Institute of Physics, Polish Academy of Sciences,
    Al Lotnik\'ow 32/46, PL-02-668 Warsaw, Poland}

\author{Piotr Wojnar}
\email{wojnar@ifpan.edu.pl}
    \affiliation{Institute of Physics, Polish Academy of Sciences,
    Al Lotnik\'ow 32/46, PL-02-668 Warsaw, Poland}

\begin{abstract}

The structure of the ground state exciton of Cd(Se,Te) quantum dots embedded in ZnTe matrix is studied experimentally using photoluminescence spectroscopy and theoretically using ${\bf k}\cdot{\bf p}$ and configuration interaction methods. The experiments reveal a considerable reduction of fine-structure splitting energy of the exciton with increase of Se content in the dots. That effect is interpreted by theoretical calculations to originate due to the transition from spatially direct (type-I) to indirect (type-II) transition between electrons and holes in the dot induced by increase of Se. The trends predicted by the theory match those of the experimental results very well.
The theory identifies that the main mechanism causing elevated fine-structure energy in particular in type-I dots is due to the multipole expansion of the exchange interaction. Moreover, the theory reveals that for Se contents in the dot $>0.3$, there exist also a {\bf peculiar type of confinement showing signatures of both type~I and type~II} and which exhibits extraordinary properties, such as almost purely light hole character of exciton and toroidal shape of hole states.
%

\end{abstract}

\maketitle

\section{Introduction}

Key components of the future quantum devices for usage in information technology will be on demand sources of single photons and entangled photon-pairs. A prominent candidate systems in this respect are currently the quantum dots (QDs). Mainly semiconductor type-I QDs where both electron and hole wavefunctions are bound inside QD body, show excellent optical properties combined with their compatibility with current semiconductor processing technology and, moreover, they offer the potential for scalability~\cite{Aharonovich2016,Senellart2017,Thomas2021,Tomm2021,Orieux2017,Huber2018a,Klenovsky_IOP2010,Klenovsky2015}. QDs currently cover a rather wide range of topics, from quantum cryptography protocols~\cite{Muller2014,Strauf2007}, sources of polarization-entangled photon pairs~\cite{Huber2018,Liu2019,Huang2021}, quantum key distribution~\cite{BassoBasset2021,Schimpf2021}, quantum gates~\cite{Stevenson2006,Burkard1999,Klenovsky2019,Klenovsky2016,Krapek2010}, or as nanomemories~\cite{Marent2011,Marent2009_microelectronics,Bimberg2011_SbQDFlash, Marent_APL2007_10y,Sala2016,Sala2018,Klenovsky2019}.

One of the key challenges for turning QDs into sources of entangled photons is to zero the tiny energy separation of the bright doublet of the ground state exciton (X$^0$), dubbed the fine-structure splitting (FSS). That can be achieved,~e.g., by externally applying elastic strain~\cite{Trotta2012,Trotta2013,Trotta2016,Martin-Sanchez2017,Aberl2017,Klenovsky2018}, electric~\cite{Huang2021}, or magnetic~\cite{Lobl2019,Huber2019,Csontosova2020} fields. Further option to have QDs with negligible FSS is provided by growing QDs on lattice matched materials~\cite{Rastelli2004,Sala2020}.

Another route of obtaining small FSS is utilizing type-II QDs where one of the quasiparticles, electron or hole, is bound outside of QD body, while the other resides inside~\cite{Matsuda2007,Miloszewski2014,Jo2012,Young2014,Klenovsky2010,Klenovsky_IOP2010,Klenovsky2017}. The aforementioned type-II QDs were mostly realized on group III--V materials, InAs, GaAs, and GaSb. However, there exist another class of type-II QD structures based on II--VI materials, like the Cd(Se,Te)/ZnTe dots. {\bf Note, however, that our purpose in this work is not to study the feasibility of Cd(Se,Te)/ZnTe dots for generation of entangled photons but rather to study the reduction of FSS with increased electron-hole spatial separation.}

CdSe/ZnTe is a semiconductor system, which is well-known for its type-II band alignment, in which the spatially indirect optical emission appears in the near infrared spectral region,~i.e.,~at 1.0~eV--1.1~eV. This fact has been demonstrated experimentally in CdSe/ZnTe quantum wells~\cite{Mourad2012}, CdSe/ZnTe core/shell nanowires~\cite{Wojnar2021} and colloidal core/shell nanocrystals~\cite{Kim2003}.
However, the lattice mismatch which is the driving force for the formation of self assembled QDs is very small and amounts to 0.003 in this material system, which prevents the formation of type-II CdSe/ZnTe QDs. 

On the other hand, a sufficiently large lattice mismatch of 0.07 is present in CdTe/ZnTe semiconductor system which has enabled the growth of CdTe/ZnTe QDs by molecular beam epitaxy~\cite{Tinjod2003,Wojnar2011}. Subsequently, their optical properties have been subject of extensive investigations~\cite{Leger2007,Kazimierczuk2011,Smolenski2015}.
In particular, it was found that this semiconductor system is characterized by the type-I confinement which is manifested by quite short excitonic lifetimes,~i.e., below 500~ps~\cite{Kazimierczuk2010}. However, the valence band offset in these structures is quite small~\cite{Deleporte1992} whereas only strain effects are responsible for its type-I character. That is why the addition of a certain amount of selenium into Cd(Se,Te) QD-layer, leading to the shift the valence band towards lower energies, results in the transition from type-I to type-II confinement~\cite{Baranowski2020}. At the same time the Cd(Se,Te)/ZnTe lattice mismatch is sufficiently large to induce the QD formation and the optical emission from these structures is intense enough to enable the observation of the emission from individual QDs, which enables the unique study of FSS in those type-II QDs that we discuss in this work.  

The paper is organized as follows. We start with description of experiments,~i.e., the growth of Cd(Se,Te)/ZnTe QDs and measurements of their optical emission, revealing FSS of that system as function of Se content. That is followed by theory discussion of electronic structure of Cd(Se,Te)/ZnTe QDs, starting from analysis of the single-particle states and carrying on to computations of correlated excitons, finally showing that the trends predicted by theory match those of the experimental results very well.
Furthermore, theory shows a rather unusual behavior of Cd(Se,Te)/ZnTe QDs related to light hole exciton and Aharonov-Bohm effect.

\section{Experiment}

\begin{figure}
	\includegraphics[width=0.45\textwidth]{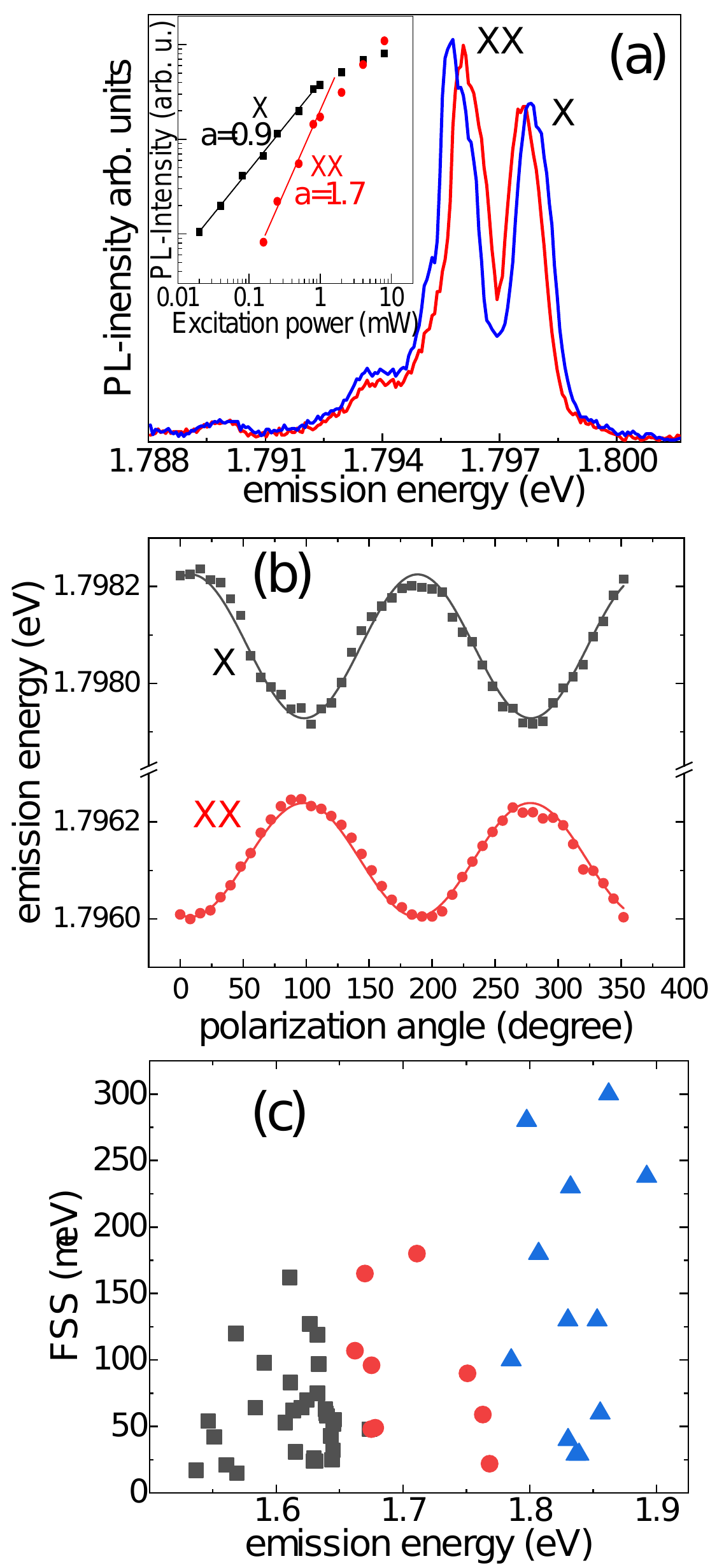}
	\caption{Fine structure splitting (FSS) of individual Cd(Se,Te)/ZnTe quantum dots (QDs) for (a) exciton (X) and biexciton (XX) emission from a single Cd(Se,Te)/ZnTe QD measured in two orthogonal linear polarizations corresponding to the anizotropy axes of the dot (b) spectral position of the exciton and biexciton emission from the same dot as a function of the polarization angle. Solid lines represent fits with sine square functions from which the FSS value of 275~$\mu$eV is determined (c) FSS values for various individual QDs as a function of the exciton emission energy. {\bf The inset of panel (a) shows the measured (points) and fitted (lines) dependencies of X (black) and XX (red) with slopes $a=0.9$ and $a=1.7$, respectively.} The dots are taken from three samples with a different average Se concentrations of~0.002,~0.03,~and~0.1 which is marked in~(c) with different colors: blue, red, and dark grey, respectively. Temperature during these measurements was kept at 7~K, the excitation laser wavelength was 405~nm and the laser spot diameter was $\sim 3 \mu$m.}
	\label{fig:uPLmeas}
\end{figure}

The samples containing self assembled Cd(Se,Te)/ZnTe QDs are grown by the molecular beam epitaxy. The details of the growth procedure are described in Ref.~\cite{Baranowski2020}. The optically active part of the samples consists of a layer of Cd(Se,Te) QDs embedded in ZnTe matrix. Three samples with a different average Se concentrations within the dots equal to~0.002,~0.03, and~0.1 are investigated for the purposes of the present work. {\bf Se content within QDs can be effectively changed by varying the growth parameters during the deposition of the QD-layer. That layer consists consecutively of three CdTe monolayers, one CdSe sub-monolayer, and two CdTe monolayers. Depending on the coverage of the central CdSe monolayer, which is controlled by the exposure time of Se-flux, the average Se concentration can be varied from~0 up to~0.17. The largest Se-concentration corresponds to the deposition of a complete central CdSe monolayer.} After the deposition of the QDs-layer the QD-formation process does not take place spontaneously despite of the large lattice mismatch, as is common for II-VI semiconductor system. It has to be additionally induced by tellurium deposition at low substrate temperature and its subsequent  thermal desorption~\cite{Wojnar2011,Tinjod2003}. In the final step, the dots are capped with a 50~nm thick ZnTe layer.

Photoluminescence (PL) measurements performed at low temperature reveal that the emission energy strongly depends on Se concentration within the Cd(Se,Te) QDs which is induced, most likely, by the change of confinement from type-I to type-II~\cite{Baranowski2020}. In particular, it is found that the maximum emission energy amounts to 1.98~eV, 1.83~eV, and 1.69~eV for the investigated mean Se concentrations of 0.002, 0.03, and 0.1, respectively.  This choice of the samples along with the inhomogeneous broadening of the emission bands, which amounts typically to 80~meV, ensures that the emission lines from individual QDs can be found in the entire spectral range, from 1.5~eV up to 1.9~eV.  

In order to assess the emission from individual QDs, $\mu$-PL measurements in which the excitation laser spot is reduced to 3~$\mu$m are performed. Further reduction of the excitation area is obtained using apertures with diameter of 400~nm within a 150~nm thick gold layer deposited on top of the structures. For the measurements, the samples are placed inside a continuous flow cryostat in which the temperature is kept at 7~K.

Several emission lines with the spectral width in the range of 500~$\mu$eV -- 1~meV  originating from individual QDs are observed. In order to determine the corresponding excitonic FSS values, linear polarization of the optical emission spectrum has been measured. This study is performed in geometry in which light propagates perpendicular to the sample surface and the linear polarization vector is always parallel to the sample plane. It is found that the emission energy slightly depends on the linear polarization angle for all measured bands. In Figure~\ref{fig:uPLmeas}~(a), the emission lines are measured in two orthogonal linear polarizations corresponding to the largest change of the emission energy. In such configuration, the energy difference is given by the value of FSS. In order to determine the best FSS-values, the spectral position is plotted as a function of the polarization angle for both emission lines, see Fig.~\ref{fig:uPLmeas}~(b). It is found that this dependence can be well fitted with a sine square function, whereas its amplitude gives us directly the FSS value~\cite{Leger2007,Kowalik2007}.
FSS values of the two emission bands presented in Fig.~\ref{fig:uPLmeas}~(a)~and~(b) are found to be very similar. However, both polarization angle dependencies are shifted by 90$^\circ$ with respect to each other, see Fig.~\ref{fig:uPLmeas}~(b). This feature is characteristic for biexciton and single exciton emission and indicates that both bands originate from the same QD. In order to identify whether the particular band corresponds to the single exciton or to biexciton emission, excitation power dependence of the optical emission spectrum has been measured,\textbf{ inset}. The intensity of the high energy line at 1.798~eV increases almost linearly with increasing excitation power whereas the intensity of the low energy line at 1.796~eV increases superlinearly which leads us to associate them to the single exciton and biexciton emission, respectively.  

In Figure~\ref{fig:uPLmeas}~(c), FSS values from over 40 individual QDs are plotted as a function of the single exciton emission energy. A large distribution of FSS values decreasing from $\sim 300$~$\mu$eV to almost zero is found among the investigated dots. Most importantly, they depend conspicuously on the exciton emission energy. Significantly, smaller FSS-values are observed, in average, for the dots emitting at lower energies compared to the dots emitting at higher energies. {\bf The large variation of these values at a fixed energy results, most likely, from the anisotropy of the potential localizing charge carriers which is induced by the shape and/or strain anisotropy of the dots similar to CdTe/ZnTe QDs without Se~\cite{Kazimierczuk2011}.}
On the other hand, the maximum emission energy depends primarily on the Se concentration within the dots~\cite{Baranowski2020}. 
At the same time, it is found that the sizes and shapes of the dots do not change significantly as a function of Se-content within the investigated concentration range as demonstrated previously by atomic force microscopy~\cite{Baranowski2020}.
Thus, it is reasonable to conclude that the increase of the average Se concentration within Cd(Se,Te) QDs leads to the overall decrease of FSS values.  A possible explanation of this effect relies on the change of the confinement of the dot/matrix interface character from type-I to type-II, leading to the increase of the electron-hole spatial separation. Furthermore, the effect of mutual compensation of electron and hole wavefunction anisotropies may result in the observed decrease of FSS-values in type-II QDs, as predicted theoretically in Ref.~\cite{Krapek2015}.

{\bf Based on the growth procedure and the optical measurements presented above we cannot draw any definite conclusion about the Se composition profile within the dots. In our considerations an uniform Se-distribution is assumed for simplicity reasons. In fact, the presence of Se- and Te-rich regions within the dot inducing additional electron-hole separation within the dots cannot be excluded. Such effects have already been studied in entirely type-I QD systems in which Cd(Se,Te) were embedded into ZnSe matrix~\cite{Sciesiek2016}.
One of the conclusions of that work was that the FSS values were even slightly increased in the presence of Se-atoms as compared to pure CdSe/ZnSe QDs and CdTe/ZnTe QDs. Since in the presently described Cd(Se,Te)/ZnTe QDs a decrease of the average FSS values takes place with an increasing Se-content, we do not expect that the local variation of Se and Te within the dots impacts significantly our results.}

\begin{figure}
	\includegraphics[width=0.45\textwidth]{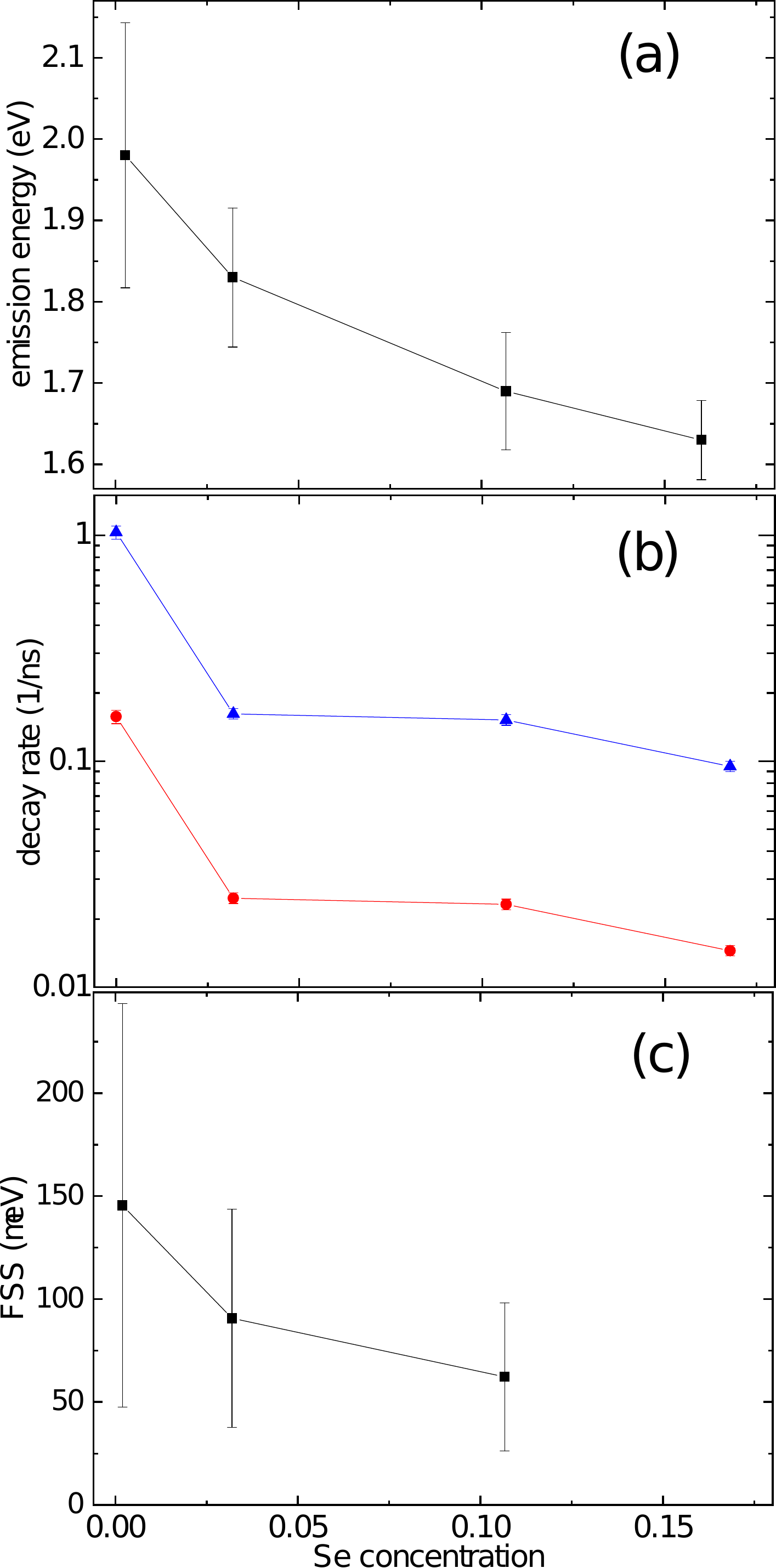}
	\caption{Dependence of selected optical properties on the Se concentration within Cd(Se,Te)/ZnTe quantum dots (a) emission energy (b) decay rates, whereas fast and slow decay are shown in blue and red, respectively, (c) average FSS determined from the data presented in Figure~\ref{fig:uPLmeas}~(c). Temperature of the measurements was 7~K and the excitation laser wavelength~405~nm.}
	\label{fig:uERateFSSmeas}
\end{figure}

The most distinct experimental trends concerning the optical emission from Cd(Se,Te)/ZnTe QDs which appear as a function of increasing Se concentration within the dots are presented in Figure~\ref{fig:uERateFSSmeas}. First of all, a considerable redshift of the emission energy from 1.98~eV down to 1.6~eV is observed,~Figure~\ref{fig:uERateFSSmeas}~(a). That is accompanied by a decrease of the decay rate by one order of magnitude, Figure~\ref{fig:uERateFSSmeas}~(b). Since the PL-decays can be well described by biexponential functions~\cite{Baranowski2020}
for all samples, the fast and slow decay rates are determined and plotted in blue and red in Fig.~\ref{fig:uERateFSSmeas}~(b), respectively. Finally, those results are compared to the dependence of FSS values as a function of Se concentration, which are the main subject of this work, see Fig.~\ref{fig:uERateFSSmeas}~(c). The values presented in Fig.~\ref{fig:uERateFSSmeas}~(c) are obtained from the arithmetic average from all QD excitons observed on a given sample. The three experimental points correspond to the three samples with different average Se concentrations investigated in this work. Despite of the fact that there is a large distribution of FSS values a clear decrease of  average FSS values with increasing Se concentration is observed. In the case of Cd(Se,Te)/ZnTe QDs with Se-content of 0.17 the optical emission was too weak to perform a detailed FSS investigation like in the other samples with lower Se-content.      

Note that the distinct decrease of the decay rates strongly indicates the type-II character of the QDs and is caused directly by the electron-hole wavefuncion spatial separation. The huge emission energy redshift of 350~meV is also consistent with the type-II confinement in CdSeTe/ZnTe QDs with relatively large Se-content. Cd(Se,Te) bandgap reduction cannot explain this effect since it is expected to amount to only 135~meV at maximum (for Se concentration of 0.4) due to the bowing effect~\cite{Wei1999}.


\section{Theory}

Based on the aforementioned experimental results we will now provide the theoretical reason for the reduction of FSS values with increasing Se content. In order to do that we calculate the correlated electronic structure of the ground state exciton (X$^0$) using a combination of the eight-band ${\bf k}\cdot{\bf p}$ method~\cite{Birner2007,t_zibold,Mittelstadt2022}, providing single-particle (SP) basis states for the configuration interaction~\cite{Schliwa:09} (CI) algorithm which we developed earlier, see Ref.~\cite{Klenovsky2017}. During the CI calculation our CI code evaluates also the emission radiative rate~\cite{Klenovsky2017,Klenovsky2019} utilizing the Fermi's golden rule~\cite{Dirac1927}.

More specifically, we consider~\cite{Klenovsky2017,Csontosova2020,Mittelstadt2022} the SP states as linear combination of $s$~orbital~like and $x$, $y$, $z$ $p$~orbital~like Bloch waves at $\Gamma$ point of the Brillouin zone,~i.e.
\begin{equation}
    \Psi_{a_i}(\mathbf{r}) = \sum_{\nu\in\{s,x,y,z\}\otimes \{\uparrow,\downarrow\}} \chi_{a_i,\nu}(\mathbf{r})u^{\Gamma}_{\nu}\,.
\end{equation}
Here $u^{\Gamma}_{\nu}$ is the Bloch wave-function of an $s$-like conduction band or a $p$-like valence band at $\Gamma$ point, $\uparrow$/$\downarrow$ mark the spin, and $\chi_{a_i,\nu}$ is~the~envelope function for $a_i \in \{ e_i, h_i \}$.

On the other hand, in CI we consider the excitonic wavefunction as a linear combination of the Slater determinants (SDs)
\begin{equation}
    \psi_i^{\rm X}(\mathbf{r}) =  \sum_{\mathit m=1}^{n_{\rm SD}} \mathit \eta_{i,m} D_m^{\rm X}(\mathbf{r}), \label{eq:CIwfSD}
\end{equation}
where $n_{\rm SD}$ is the number of SDs $D_m^{\rm X}(\mathbf{r})$, and $\eta_{i,m}$ is the $i$-th CI coefficient which is found along with the eigenenergy using the variational method by solving the Schr\"{o}dinger equation 
\begin{equation}
\label{CISchrEq}
\hat{H}^{\rm{X}} \psi_i^{\rm X}(\mathbf{r}) = E_i^{\rm{X}} \psi_i^{\rm X}(\mathbf{r}),
\end{equation}
where $E_i^{\rm{X}}$ is the $i$-th eigenenergy of excitonic state $\psi_i^{\rm X}(\mathbf{r})$, and~$\hat{H}^{\rm{X}}$ is the CI Hamiltonian which reads

\begin{equation}
\label{CIHamiltonian}
\hat{H}^{\rm{X}}=\hat{H}_0^{\rm{SP}}+\hat{V}^{\rm{X}},
\end{equation}
where $\hat{H}_0^{SP}$ and $\hat{V}^{\rm{X}}$ represent the Hamiltonian of the noninteracting SP states and the~Coulomb interaction between them, respectively. The matrix element of $\hat{V}^{\rm{X}}$ is~\cite{Klenovsky2017,Klenovsky2019,Csontosova2020}
\begin{equation}
\begin{split}
    &\bra{D_n^{\rm X}}\hat{V}^{\rm{X}}\ket{D_m^{\rm X}} = -\frac{1}{4\pi\epsilon_0} \sum_{ijkl} \iint {\rm d}\mathbf{r} {\rm d}\mathbf{r}^{\prime} \frac{e^2}{\epsilon(\mathbf{r},\mathbf{r}^{\prime})|\mathbf{r}-\mathbf{r}^{\prime}|} \\
    &\times \{ \Psi^*_i(\mathbf{r})\Psi^*_j(\mathbf{r}^{\prime})\Psi_k(\mathbf{r})\Psi_l(\mathbf{r}^{\prime}) - \Psi^*_i(\mathbf{r})\Psi^*_j(\mathbf{r}^{\prime})\Psi_l(\mathbf{r})\Psi_k(\mathbf{r}^{\prime})\}.
\end{split}
\label{eq:CoulombMatrElem}
\end{equation}
where $e$ labels the elementary charge and $\epsilon(\mathbf{r},\mathbf{r}^{\prime})$ is the spatially dependent dielectric function. Note that minus sign in front of the integral in Eq.~\eqref{eq:CoulombMatrElem} results from different sign of the charge of the electron and hole from which exciton is composed. The sixfold integral in Eq.~\eqref{eq:CoulombMatrElem} is evaluated using the~Green's function method~\cite{Schliwa:09,Stier2000,Klenovsky2017,Csontosova2020}. {\bf Note, that for $\epsilon(\mathbf{r},\mathbf{r}^{\prime})$ in Eq.~\eqref{eq:CoulombMatrElem} we use the positionally dependent bulk dielectric constant in our CI calculations.} Further, the multipole expansion of the exchange interaction is included in our CI for CI basis consisting of two electron and two hole SP ground states following the theory outlined in Refs.~\cite{Takagahara2000,Krapek2015}.


%
%
\begin{figure}[!ht]
	\includegraphics[width=0.45\textwidth]{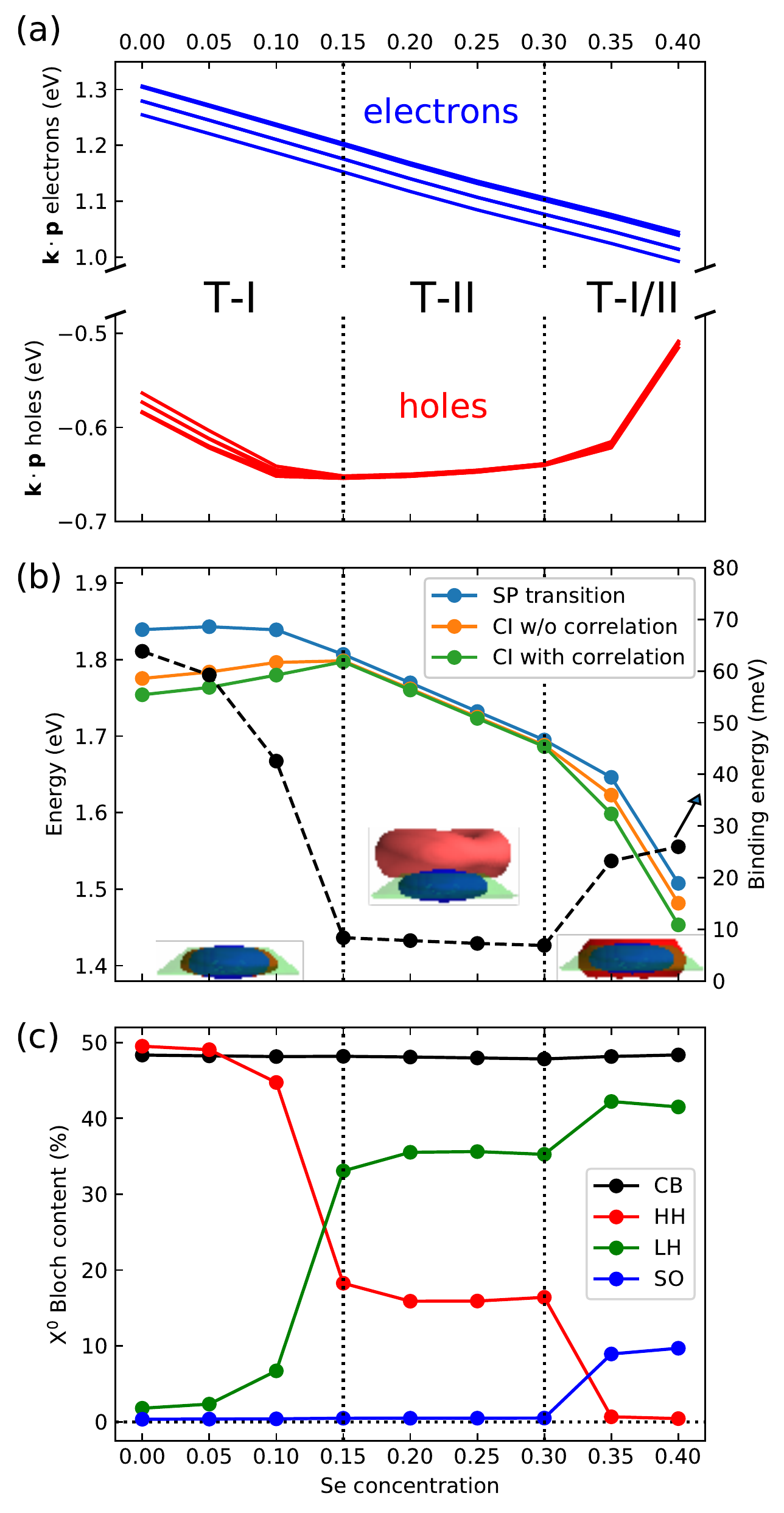}
	\caption{Electronic states of Cd(Se,Te)/ZnTe QDs. In panel (a) we show twelve single-particle (SP) energies of electrons (blue) and holes (red). The inset in (a) gives markings of different types of confinement in Cd(Se, Te)/ZnTe QDs,~i.e., type-I (mark T-I) for Se=0--0.15, type-II (mark T-II) for Se=0.15--0.3, and a {\bf type I/II} (mark T-I/II) for Se=0.3--0.4. In (b) ve show the SP energies (blue curve) and that computed using CI without (orange curve) and with (green curve) the inclusion of the effect of Coulomb correlation. {\bf In (b) we also give the binding energy of X$^0$ with respect to SP transition (black broken curve) with energy axis on the right.} The insets in~(b) show side cuts of our QD (green object) and the SP electron (blue object) and hole (red object) probability densities. Panel (c) gives the conduction (CB, black), heavy-hole (HH, red), light-hole (LH, green), and spin-orbit split off (SO, blue) Bloch band content of X$^0$ as a function of Se concentration computed~\cite{Csontosova2020} using CI with twelve electron and twelve hole SP basis states,~i.e., including the effect of correlation.
	The transitions between different confinements are marked in all panels by black dotted vertical lines.}
	\label{fig:TeorSPCIBloch}
\end{figure}

\onecolumngrid

\begin{figure}[!ht]
	\centering
	\includegraphics[width=0.95\textwidth]{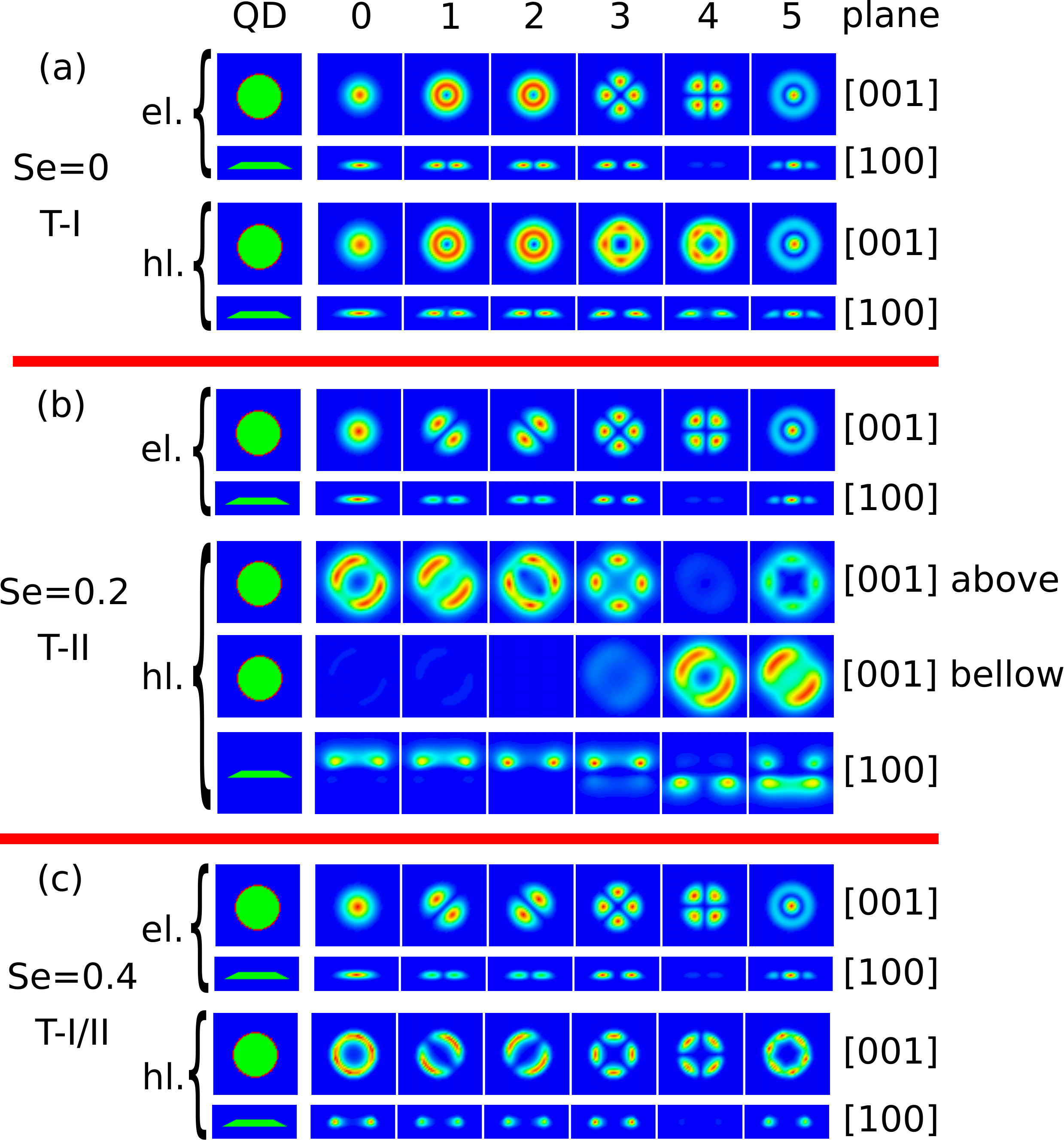}
	\caption{Cuts of the single-particle (SP) probability densities of Cd(Se,Te)/ZnTe QDs for (a) zero Se content, (b) Se content of 0.2, and (c) Se content of 0.4, corresponding to type-I, type-II, and {\bf type-I/II confinement}, respectively. The letters QD in top row mark that the first column showing the cuts of the simulated QD body, the numbers in the first row enumerate SP states, starting from the ground state marked by zero. The last column gives the Miller indices of the planes where the cut was performed in each row of the figure. The abbreviations ``el." and ``hl." mark the electrons and holes, respectively. In (b) the designations ``[001] above" and ``[001] bellow" identify that the cuts of the hole densities were performed above and bellow QD body, respectively, and correspond to the side cut given in the last row of panel (b).}
	\label{fig:ProbabDen}
\end{figure}

\twocolumngrid

\clearpage

\section{Results and Discussions}
The electronic states of Cd(Se,Te)/ZnTe QDs computed using the aforementioned ${\bf k}\cdot{\bf p}$+CI method are shown in Fig.~\ref{fig:TeorSPCIBloch}. Motivated by typical structure pinpointed in Ref.~\cite{Baranowski2020}, the computed shape of the Cd(Se,Te) QD was truncated cone with lower and upper basis diameters of 36~nm and 22~nm, respectively, and with height of 4~nm. Except of QD body, the rest of the simulation space consisted of ZnTe. After definition of the structure, the elastic strain tensor was obtained in the whole simulated structure by grid-point-wise minimization of elastic energy. Thereafter, the Coulomb potential energy, including the effects of the piezoelectricity, was obtained by solving the Poisson's equation in the whole structure. The resulting Hamiltonian matrix was then diagonalized using Nextnano++~\cite{Birner:07,t_zibold} simulation tool, which was also used for the aforementioned computation steps. {\bf Note, that all the material parameters including the effective masses were taken from the library of the Nextnano++ software~\cite{Birner:07}.}

The resulting eigenenergies and eigenfunctions are shown in Fig.~\ref{fig:TeorSPCIBloch}~(a) and Fig.~\ref{fig:ProbabDen}, respectively, for twelve electron and twelve hole SP states. Depending on Se content, we have identified three types of confinement in our QDs,~i.e., type-I for Se=0--0.15, type-II for Se=0.15--0.3, and {\bf so-called type~I/II} for Se=0.3--0.4. Note, that the type-II confinement was identified by the spatial location of hole wavefunctions \{inset in Fig.~\ref{fig:TeorSPCIBloch}~(b) and Fig.~\ref{fig:ProbabDen}~(b)\} being outside of QD body, while electrons are firmly bound inside QD for all Se contents. {\bf The type~I/II confinement is peculiar and shows features of the remaining two types of confinement, see also in the following.} 

While the energy of electron SP states reduces with similar rate for all Se contents, including that for SP excited states, the holes are affected by Se content and the associated type of confinement considerably more. While for type~I and associated Se contents $<0.15$, hole SP energy decreases (i.e., the absolute value of hole energy increases), for type~II that increases only slightly, and, finally, for {\bf type~I/II} the increase of hole SP energy (decrease of the absolute value of that) with increasing Se content is observed. 

As a result of the aforementioned discussion, the SP electron-hole transition energy \{see blue curve in Fig.~\ref{fig:TeorSPCIBloch}~(b)\} remains almost constant for increasing Se content in type~I, while its magnitude is reduced with increase of Se for type~II and type~I/II.

The Se-content-dependent energies of X$^0$ computed by CI without and with considering the effect of correlation are shown by orange and green curves, respectively, in Fig.~\ref{fig:TeorSPCIBloch}~(b). {\bf Note, that by the ``effect of correlation" we mean specifically the expansion of CI complexes into the basis consisting not only from ground but also excited SP states.} Now, the binding energy of X$^0$ {\bf \{Fig.~\ref{fig:TeorSPCIBloch}~(b)\} } compared to electron-hole transition SP energy decreases in type~I from 70~meV for Se content of zero to 10~meV for content of 0.15. The large binding energies in type~I are due to the attractive Coulomb interaction between tightly quantum confined electrons and holes in QD~\cite{Klenovsky2021}. For Se content between 0.15 and 0.3 (type II) the binding energy remains constant at 10~meV and further increases with Se content to 30~meV in type-I/II. Correlation further increases the binding energy by 20~meV for Se content of zero up to by almost 30~meV for Se content of 0.4. In type~II the additional binding energy due to correlation is $\sim2$~meV. Finally, note that the magnitude of the reduction of X$^0$ energy with increasing Se content matches that observed from spectral shift of the maximum of photoluminescence spectra in Ref.~\cite{Baranowski2020}.

In Fig.~\ref{fig:TeorSPCIBloch}~(c) we show the Se-dependent Bloch state content of X$^0$, obtained from the Bloch state composition of electron and hole SP states, utilizing the squares of CI wavefuntion coefficients $\eta_{i,m}$ from Eq.~\eqref{eq:CIwfSD}. The method was previously developed in Refs.~\cite{Csontosova2020,Huang2021}. Note, that the studied Bloch state composition in this work is that of the conduction band~(CB), heavy-hole~(HH), light-hole~(LH), and spin-orbit~(SO) valence bands. 

The content of CB in X$^0$ is $\sim50$\,\% for all studied Se contents. While HH content is also $\sim50$\,\% for smaller Se concentrations, the increase of the amount of Se causes considerable progressive admixing of LH states in X$^0$, reaching values of $\sim40$\,\% for Se contents $>0.35$. At the same time, we observe increase of admixing of SO state for Se contents $>0.3$, reaching values as high as $\sim10$\,\%. We note that for Se $<0.3$ the SO content of X$^0$ is negligibly small.

Thus, both the type-II regime and in particular {\bf type-I/II} show unusual composition of X$^0$. Strikingly, in type~I/II X$^0$ has almost LH like character with small addition of SO states and negligible HH content. Such LH X$^0$ would be advantageous in quantum information technology, {\bf such as,~e.g., enabling coherent conversion of photons into electron spins~\cite{Vrijen2001},} and was first experimentally reported in Ref.~\cite{Huo:NatPhys} for GaAs/AlGaAs QD system where the LH character was obtained by externally applied tensile strain. However, we predict that in Cd(Se,Te)/ZnTe QDs such an LH exciton is present for Se contents $>0.35$ without the necessity of any external tuning. 
%
%
%

\begin{figure}[!ht]
	\centering
	\includegraphics[width=0.45\textwidth]{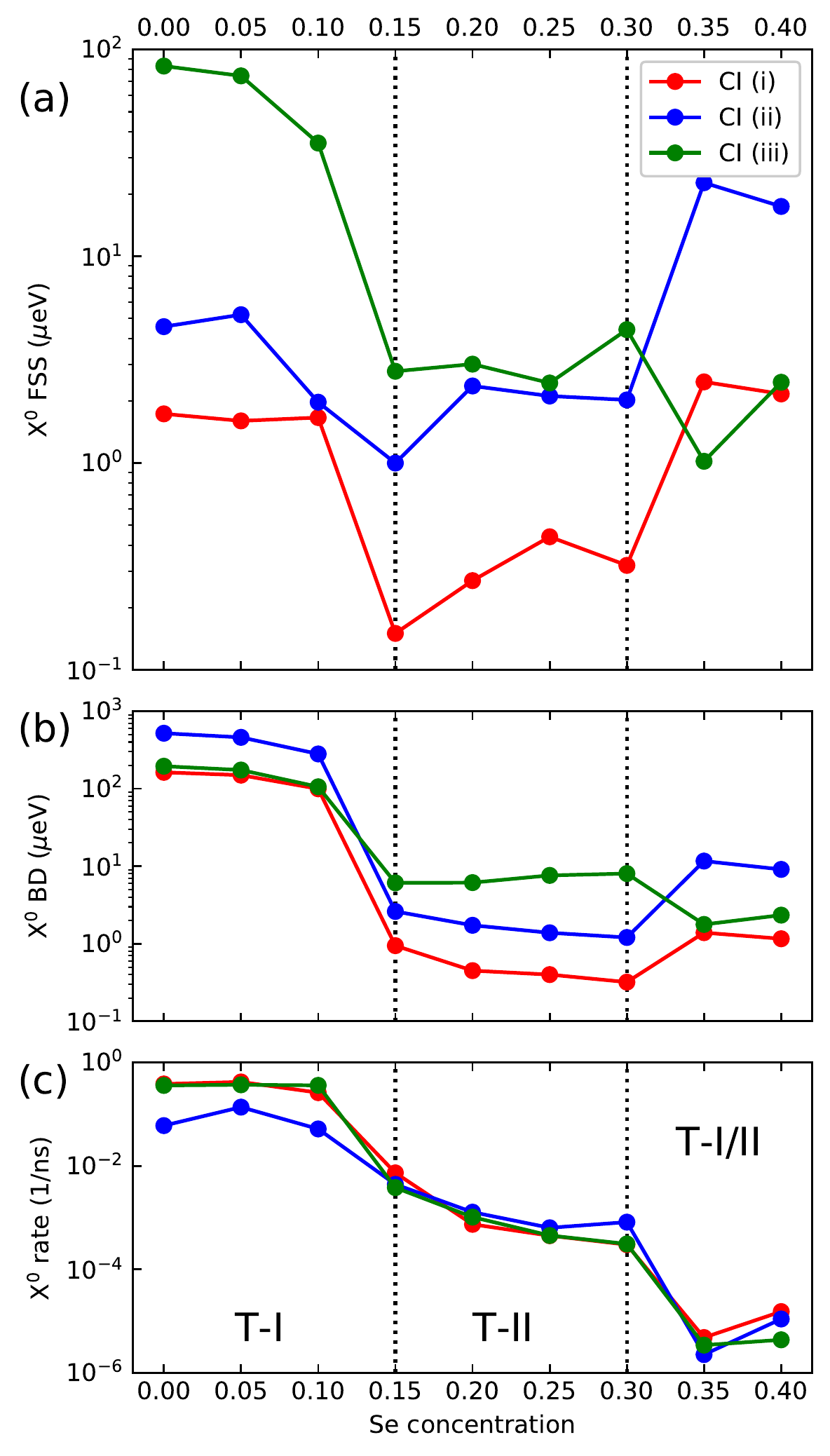}
	\caption{Fine energy structure of ground state exciton (X$^0$) in Cd(Se,Te)/ZnTe QDs as a function of Se content. In (a) we show the theory values of FSS obtained using CI with different level of approximation,~i.e., without including the effect of correlation as well as without considering the multipole expansion of exchange interaction \{``CI (i)", red curve\}, that for data with correlation included and without multipole \{``CI (ii)", blue curve\}, and without the effect of correlation but with multipole expansion included \{``CI (iii)", green curve\}. Panel (b) shows the calculated values of energy splitting between bright and dark exciton (X$^0$ BD) and in (c) we give the computed values of corresponding X$^0$ emission rates. Note that colors of theory curves in (b) and (c) correspond to the same CI approximations as was described for (a). The transitions between different confinements are marked in all panels by black dotted vertical lines.}
	\label{fig:FSSBDteorVsMeas}
\end{figure}

We now turn our attention to the theory analysis of the fine-structure of X$^0$ and show the results of that in Fig.~\ref{fig:FSSBDteorVsMeas}. We computed the fine structure employing three levels of approximation in CI,~i.e., (i) without considering the effect of correlation and with monopole-monopole term of the exchange interaction only, (ii) with correlation and monopole-monopole term of exchange, and (iii) without correlation but with assuming the monopole-monopole, monopole-dipole, and dipole-dipole terms of the exchange interaction what we call multipole expansion.~\cite{Krapek2015,Krapek2016}

The results for FSS of X$^0$ are given in Fig.~\ref{fig:FSSBDteorVsMeas}~(a) by circles and full curves.
Firstly, we see that FSS values computed using approximations (i) and (ii) defined in previous paragraph are $<10\,\,\mathrm{\mu eV}$ for all studied Se concentrations. However, the multipole expansion of exchange,~i.e., point (iii) above, gives more realistic estimation of FSS, when compared to experimental data \{Figs.~\ref{fig:uPLmeas}~and~\ref{fig:uERateFSSmeas}\}, in particular for type-I confinement. 
We, thus, conclude that FSS in type~I and type~II confinement of our system is dominated by multipole expansion of the exchange interaction, predominantly by the dipole-dipole term~\cite{Klenovsky2021}. However, in the regime of {\bf type~I/II} FSS is increased up to $20\,\,\mathrm{\mu eV}$ due to the effect of Coulomb correlation.

In Fig.~\ref{fig:FSSBDteorVsMeas}~(b) we show the energy difference between optically active (bright), and inactive (dark) X$^0$ doublets and we mark that as X$^0$~BD in that panel. We see that similarly to FSS, X$^0$~BD is larger in type-I regime. Interestingly, contrary to FSS, BD is dominated in type~I by the approximation (ii),~i.e., the correlation, which causes FSS to be more than twice larger than that found by approximations~(i)~and~(iii). On the other hand, in type~II, BD is caused predominantly by approximation (iii),~i.e., multipole expansion. Finally, BD for {\bf type~I/II} is again caused mainly by correlation \{approximation (ii)\}.

Furthermore, we see the theory prediction of X$^0$ radiative rate, computed by our ${\bf k}\cdot{\bf p}$+CI computational complex, in Fig.~\ref{fig:FSSBDteorVsMeas}~(c). 
We observe that the computed rates for all three approximations (full curves) are close to experimental results in Fig.~\ref{fig:uERateFSSmeas}~(b). As expected, the emission rate of our dots in type~II ($\sim 10^{-3}$~1/ns) is reduced by two orders of magnitude compared to type~I ($\sim 10^{-1}$~1/ns). Moreover, for the {\bf type-I/II} regime that is reduced by another two orders of magnitude (to $\sim 10^{-5}$~1/ns). {\bf However, we note that in our CI calculations we omit the short-range interaction within the unit crystal cell. Thus, we neglect also the effect of that on the emission rate, resulting in the difference to the experimental data.}

{\bf The aforementioned behavior can be understood with inspection of the probability densities given in Fig.~\ref{fig:ProbabDen}. We see that in type I \{Fig.~\ref{fig:ProbabDen}~(a)\} both electron and hole probability densities reside inside QD body, hence, the overlap is largest of the studied types of confinement. On the other hand, in type II \{Fig.~\ref{fig:ProbabDen}~(b)\} while the electrons still reside inside QD, holes are pushed to the surrounding ZnTe material above and below QD, what also explains the reduced oscillator strength of X$^0$. The confinement for holes is provided by the bandedge changes caused by the elastic strain around QD and piezoelectricity originating in lattice mismatch between dot and buffer materials~\cite{Klenovsky2010,Klenovsky_IOP2010}. Finally, for type~I/II in Fig.~\ref{fig:ProbabDen}~(c) we find the holes to be confined again inside QD body. However, holes, in particular those lowest in energy, are confined on the outskirts of QD and their overlap with electrons is very faint as the wavefunctions almost exactly ``miss" each other. It is that kind of behavior, combining localization of holes inside QD with very small overlap with electrons, that led us to the nomenclature of this confinement as type-I/II.
}

However, the aforementioned analysis of the radiative emission rate of X$^0$ from Cd(Se,Te)/ZnTe QDs indicates that both type~II and {\bf type~I/II} regimes are very hard to be accessed by optical means, {\bf as the dots would emit one photon in $\sim 1\,\mu{\rm s}$ or even in $\sim 100\,\mu{\rm s}$ for the former and latter confinements, respectively.}

Finally, we would like to comment on the topology of the hole wavefunctions in Cd(Se,Te)/ZnTe QDs. From the probability densities shown in Fig.~\ref{fig:ProbabDen} we can observe, that the holes for type-II and {\bf type-I/II} \{Fig.~\ref{fig:ProbabDen}~(b)~and~(c)\} have toroidal shape for both ground and excited SP states. Thus, the hole states in Cd(Se,Te)/ZnTe QDs with type-II or type-I/II might be utilized for the realization of the Aharonov-Bohm effect~\cite{Aharonov1959}, similarly as was proposed,~e.g., in Ref.~\cite{llorens_topology_2019}.

\section{Conclusions}

We have studied the excitonic structure of Cd(Se,Te) QDs embedded in ZnTe matrix. The photoluminescence spectroscopy analysis revealed reduction of FSS of ground state exciton with increasing Se content in dots. This was found to be associated with type-II character of the dots for larger Se contents. We confirmed that by detailed ${\bf k}\cdot{\bf p}$ and CI calculations, the results of which explained the experimentally observed trends very well.
The theory identified the main mechanism causing larger FSS, in particular in type-I dots, to be due to the multipole expansion of the exchange interaction. Furthermore, using our theory we found that for Se contents in the dot larger than $\sim0.3$ a {\bf peculiar type~I/II} confinement occurs, causing an almost purely light hole character of exciton and toroidal shape of hole states.
%

\section{Acknowledgments}

P.W. acknowledges the financial supported from the  National Centre of Science (Poland) through grant 2017/26/E/ST3/00253. 

P.K. was financed by the project CUSPIDOR, which has received funding from the QuantERA ERA-NET Cofund in Quantum Technologies implemented within the European Union's Horizon 2020 Programme. In addition, this project has received national funding from the Ministry of Education, Youth and Sports of the Czech Republic and funding from European Union's Horizon 2020 (2014-2020) research and innovation framework programme under Grant agreement No. 731473.
The work reported in this paper also was partially funded by projects 20IND05 QADeT, 20FUN05 SEQUME, 17FUN06 SIQUST that received funding from the EMPIR programme co-financed by the Participating States and from the European Union's Horizon 2020 research and innovation programme.

\clearpage


\newcommand{\noopsort}[1]{} \newcommand{\printfirst}[2]{#1}
\newcommand{\singleletter}[1]{#1} \newcommand{\switchargs}[2]{#2#1}
%

\end{document}